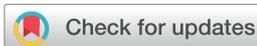

Cite this: *Soft Matter*, 2024, **20**, 9312

# Magnetic colloidal single particles and dumbbells on a tilted washboard moiré pattern in a precessing external field†

Farzaneh Farrokhzad, [a] Nico C. X. Stuhlmüller, [d] Piotr Kuświk, [b] Maciej Urbaniak, [b] Feliks Stobiecki, [b] Sapida Akhundzada, [c] Arno Ehresmann, [c] Daniel de las Heras [d] and Thomas M. Fischer [*a]

We measure the dynamical behavior of colloidal singlets and dumbbells on an inclined magnetic moiré pattern, subject to a precessing external homogeneous magnetic field. At low external field strength single colloidal particles and dumbbells move everywhere on the pattern: at stronger external field strengths colloidal singlets and dumbbells are localized in generic locations. There are however nongeneric locations of flat channels that cross the moiré Wigner Seitz cell. In the flat channels we find gravitational driven translational and non-translational dynamic phase behavior of the colloidal singlets and dumbbells depending on the external field strength and the precession angle of the external homogeneous magnetic field.

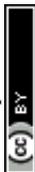

## 1. Introduction

Current research focuses on the electric conductivity and transport behavior in twisted hexagonal structures such as twisted bilayer graphene,[1–7] and twisted bilayers of the transition metal dichalcogenide family[8] because of their non-conventional superconductive[9–15] and ferromagnetic[16,17] phase behavior. Information on the physics of these systems can be gained by investigating other twisted systems: The transport of waves in twisted photonic[18–22] or acoustic[23–27] crystals as well as in vortex lattices[28] is based on similar topological transport behaviour.

We have focussed on the transport properties of classical macroscopic magnetic particle systems[29,30] as well as on transport properties of soft matter magnetic colloidal particle systems[31–40] and specifically subject to twisted magnetic patterns.[41–43] The magnetic potential of twisted patterns subject to an external drift force is a special form of a tilted washboard potential.[44–49] Tilted washboard potentials show interesting transitions[50] in their transport properties as a function of their tilt.

Here, we study the motion of single colloidal particles and of colloidal dumbbells above inclined flat channels[41] created by a magnetic moiré pattern being an overlay of two magically twisted hexagonal (or square) generator patterns of alternating magnetization (see Fig. 1). The resulting magnetic moiré pattern creates a magnetic field $H_p$ that is periodic with a hexagonal (square) shaped moiré Wigner Seitz cell. Superposition of an external magnetic field that is much stronger than the pattern field leads to a colloidal potential that consists of mostly localized potential minima and maxima. There are however extended regions of negative interference within the superposition where the potential is almost flat, called the flat channel. A flat colloidal potential channel follows a zig-zag path through the moiré Wigner Seitz cell. The corrugation of the potential above the flat channel is significantly weaker than the modulation of the potential between the localized maxima and minima. We immerse paramagnetic colloidal particles in water above the pattern and let them sediment to an equilibrium position a few nanometers above the pattern. There, they either rest on the pattern as single colloidal particles or self assemble to colloidal dumbbells of two particles held together *via* dipolar interactions.[36–38] When the external magnetic field is not normal to the pattern the flat channel direction and the external magnetic field compete to orient the colloidal dumbbells.

We apply a time dependent homogeneous external magnetic field precessing around the pattern normal at a fixed precession angle. When we sufficiently incline the moiré pattern gravity can drive the colloidal particles and colloidal dumbbells through the flat channels, while colloids in the localized minima always remain immobile. Due to the competition of anisotropic interactions, we distinguish two forms of motion along the flat channels. We find dumbbells slithering along the flat channel with the head of the colloidal dumbbell always pointing in direction of the channel and we find precessing colloidal dumbbells sliding along the flat channel and at the same time precessing with the external field. We present a dynamic phase diagram of the various transport modes in both hexagonal twisted and square twisted patterns.

[a] *Experimentalphysik X, Physikalisches Institut, Universität Bayreuth, D-95440 Bayreuth, Germany. E-mail: thomas.fischer@uni-bayreuth.de*
[b] *Institute of Molecular Physics, Polish Academy of Sciences, 60-179 Poznań, Poland*
[c] *Institute of Physics and Center for Interdisciplinary Nanostructure Science and Technology (CINSaT), University of Kassel, D-34132 Kassel, Germany*
[d] *Theoretische Physik II, Physikalisches Institut, Universität Bayreuth, D-95440 Bayreuth, Germany*

† Electronic supplementary information (ESI) available. See DOI: https://doi.org/10.1039/d4sm01183j

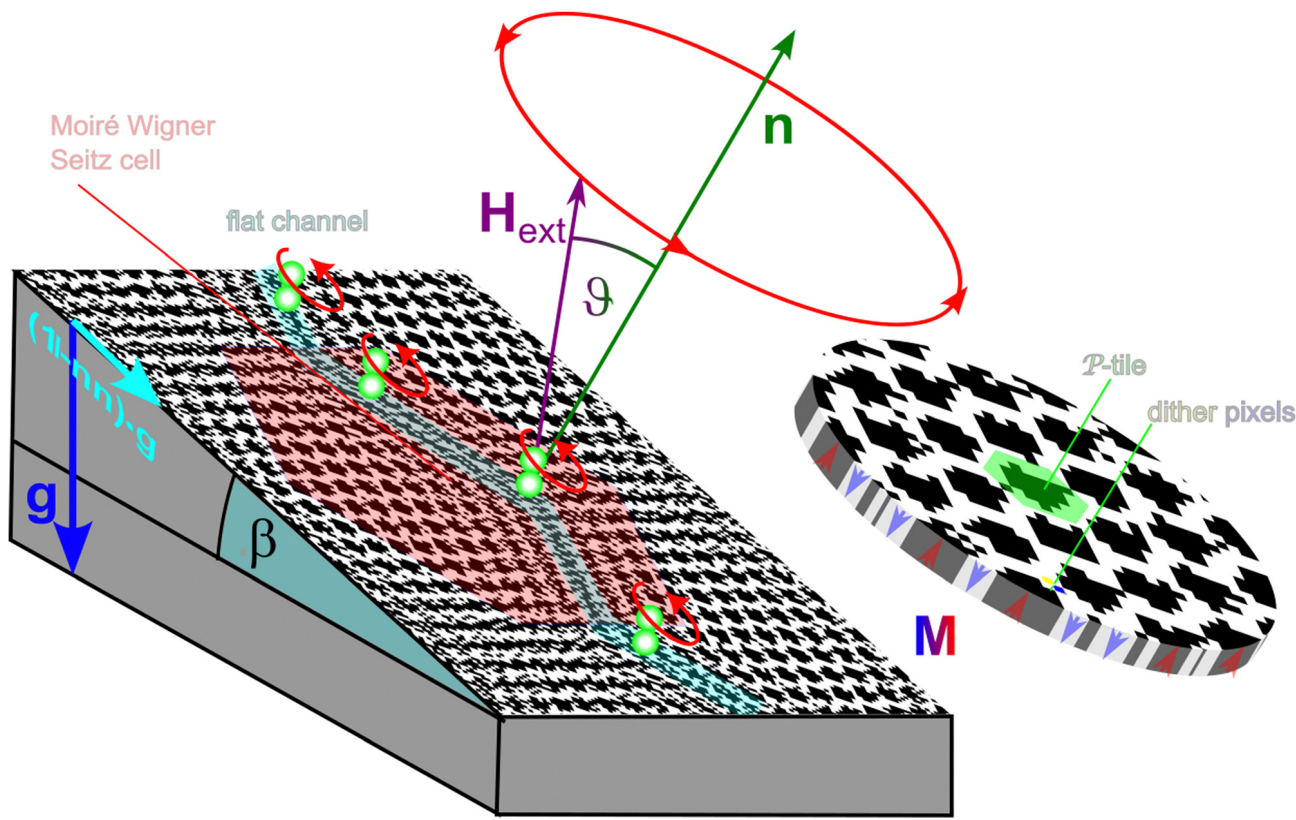

Fig. 1 Scheme of the setup: a magnetic moiré pattern with (black) up and (white) down magnetized dithered domains mathematically computed from two superposed magically twisted periodic magnetic generator patterns and then imprinted. The moiré pattern is a pattern of three different length scales. The moiré pattern is periodic with a hexagonal moiré Wigner Seitz cell. We can dissect the moiré Wigner Seitz cell into smaller hexagonal $\mathcal{P}$-tiles (see the green $\mathcal{P}$-tile in the magnified disk-shaped subregion; the unit vectors of the $\mathcal{P}$-tile are orthogonal to the unit vectors of the moiré Wigner Seitz cell). The generic $\mathcal{P}$-tile contains a localized minimum and maximum of the colloidal potential. The smallest length scale of the pattern is introduced by the dithering procedure (In the diskshaped subregion one white and one black dither pixel are recolored in yellow and blue). There are non generic $\mathcal{P}$-tiles that connect to form a flat channel (cyan) following a zig-zag path through the moiré Wigner Seitz cell. Due to the dithering, the particle-surface potential above the flat channel is corrugated and rough on the scale of the dithers. The moiré pattern is inclined with respect to the direction of gravity $\hat{\mathbf{g}}\cdot\mathbf{n} = -\cos\beta$ and subject to a precessing homogeneous (angular frequency $\omega$) external field with precession angle $\mathbf{H}_{ext}\cdot\mathbf{n} = \cos\vartheta$. Paramagnetic colloidal particles sediment onto the pattern and form colloidal dumbbells via dipolar interaction. While colloidal dumbbells in generic $\mathcal{P}$-tiles cannot move, colloidal assemblies (some singlets, mainly dumbbells) that reside within the non-generic flat channels may or may not slide down along them exhibiting different forms of orientational dynamics.

## 2. Results

We use a magnetic Co/Au multilayer, which has been patterned by keV He$^+$-ion bombardment through a lithographical mask[51,52] in a home-built bombardment stage.[53] Instead of creating a magnetic moiré pattern by twisting two thin film patterns, we lithographically produce the moiré pattern by calculating the magnetization due to superposition of the two generator patterns and imprinting the corresponding magnetization directly into one magnetic thin film. The quasi two dimensional magnetization of our moiré pattern in this single film of thickness $t$ = 5 nm reads[41]

$$M = M_s \mathcal{D}\left(\sum_{p=\pm}\sum_{i=1}^{2n}\left[\cos(\mathbf{k}^i \cdot \mathbf{s}_p(\mathbf{r})) + t_n\right]\right) \quad (1)$$

with $M_s \approx$ 1420 kA m$^{-1}$ the saturation magnetization of Co. The normal component of the pattern magnetic field satisfies the thin film boundary condition $H_z^p = ktM$ right at the film surface. $\mathcal{D}$ denotes a dithering procedure that converts a continuous gray scale image into a dithered image having only the digitized values $\pm 1$ and with pixel size of 1 µm for the square ($n$ = 2) and 2 µm for the hexagonal ($n$ = 3) pattern. The first sum runs over $p$ = + and $p$ = −. Each term creates one of two generator patterns with a generator Wigner Seitz cell of hexagonal ($n$ = 3) or square ($n$ = 2) symmetry and generator lattice constant $a$ = 14 µm. The $\mathbf{k}^i = \mathbf{R}_{\pi/n}^{i-1}\cdot\mathbf{k}^1$ in the first term of eqn (1) are the $2n$ ($i$ = 1,...,2$n$) primitive reciprocal unit vectors of magnitude $k = 2\pi/a \sin(\pi/n)$ of the non rotated hexagonal ($n$ = 3) or square ($n = 2$) generator patterns. The matrices $R_{\pi/n}$ are rotation matrices by the angles $\pi/n$ generating a pattern of the appropriate rotation symmetry. The shift vectors $s_\pm(r) = R_{\pm\alpha/2}^{-1} \cdot (r - r_{\text{center},\pm}) \stackrel{\text{mod} a_1, a_2}{=} s_\pm(0) + R_{\pm\alpha/2}^{-1} \cdot r$ of both patterns are the vectors from the nearest generator Wigner Seitz cell centers in each of the rotated generator patterns toward the lateral 2D-position of interest $r$, but rotated back into the unrotated generator pattern orientation. Non generic transport behavior is predicted for magic twist angles in smooth twisted colloidal systems.[41] This non generic behavior disappears in magically twisted system including disorder.[43] The $R_{\pm\alpha/2}$ are rotation matrices by $\pm\alpha/2$ which is half of a magic twist angle $\alpha_k^n = 2\arctan[\sin(\pi/n)/(nk + 1 + \cos(\pi/n))]$. We use $\alpha_7^3 = 4.40846°$ for the hexagonal and $\alpha_{13}^2 = 4.24219°$ for the square pattern. The choice of magic twist angle ensures a minimal size of the final moiré Wigner Seitz cell (with magically twisted moiré unit vectors $a_i^{\mathcal{TW}} = [\sin(\pi/n)/n \sin(\alpha/2)]\{a_i + a_{i+1}\}$). We use shift vectors $s_+(0) = 0$ and $s_-(0) = a_1/2$ that centers the resulting flat channel of the colloidal potential to the origin of the moiré Wigner Seitz cell. The abbreviation (mod $a_1, a_2$) above the equal sign indicates that the left and right side of the equation are equal up to differences of integer multiples of the primitive generator lattice vectors of one of the unrotated generator patterns. The parameter $t_n$ in (1) is chosen such that the average magnetization of the moiré pattern vanishes. An example of the dithered twisted hexagonal moiré pattern is shown in Fig. 1 with black up- and white down- magnetized domains. We can dissect the moiré Wigner Seitz cell into $\mathcal{P}$-tiles (with primitive tile vectors $a_i^{\mathcal{P}}$)[43] that are a little bit larger ($a_i^{\mathcal{P}} = a_i/2\cos(\alpha/2)$) than one quarter of the generator unit cells (with primitive generator unit vectors $a_i$). The generic hexagonal $\mathcal{P}$-tile harbors one minimum (one black domain) and two maxima (two white extended domain vertices) of the colloidal potential. The generic square $\mathcal{P}$-tile harbors one minimum (one black domain) and one maximum (one white domain) of the colloidal potential. In both hexagonal and square twisted patterns, non-generic $\mathcal{P}$-tiles connect to form a flat channel the corrugation of which is much smaller than the modulation within a generic $\mathcal{P}$-tile. Like other zig-zag paths in colloidal science[54] or in dissipative physics[55] our connected flat channel is chiral. Corners along the zig zag path are generically different from each other for non magic angles, but become equivalent for magic angles.[41] This is the reason why in smooth twisted potentials the transport behavior is very different at magic angles as compared to the generic non-magic twist angles.[43]

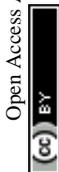
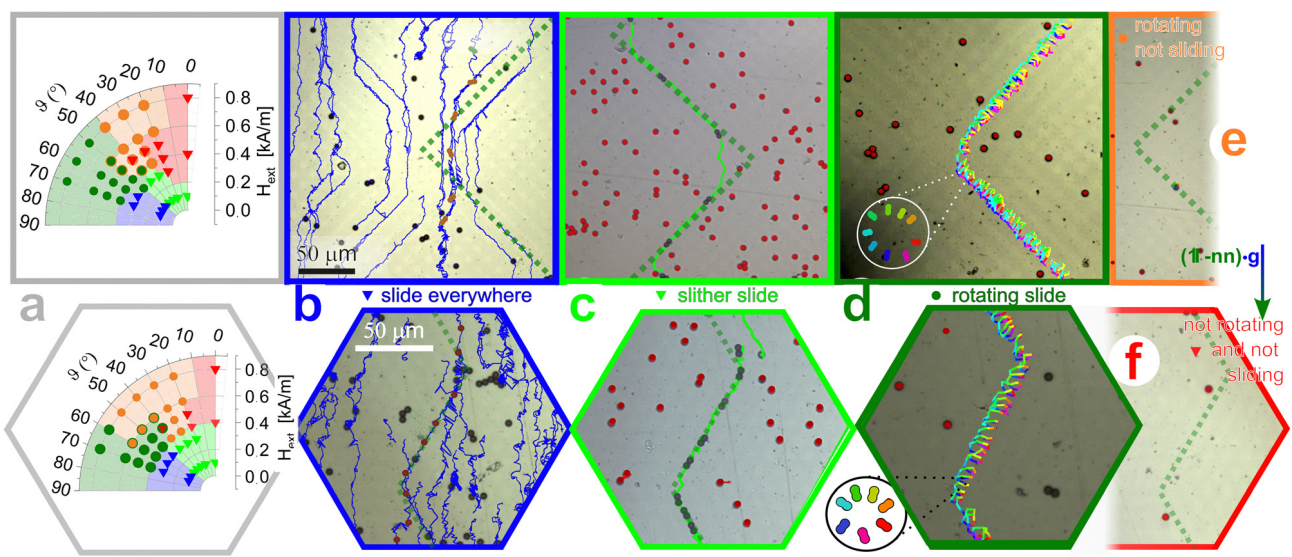

Fig. 2 Transport modes (a) dynamical phase diagrams (polar plot) of the inclined dithered twisted square (top) and hexagonal (bottom) pattern as a function of the precession angle $\vartheta$ and the strength of the external magnetic field $H_{\text{ext}}$. Symbols are experimentally measured data points. The color indicates the observed transport mode of the colloidal singlets and dumbbells. The color coded areas in the diagram are a guide to the eye. (b) Overlay of microscope images of the sliding motion in generic and non generic $\mathcal{P}$-tiles of the inclined dithered twisted square (top) and hexagonal (bottom) pattern. The in-plane direction of gravity $(1 - nn) \cdot g$ in all panels (a)–(f) is from the top to the bottom of the moiré Wigner Seitz cell. The colloidal particles and dumbbells move everywhere. The directions of motion is at an angle $\pm\pi/2n$ with respect to the inclination $(1 - nn) \cdot g$ direction or along the inclination direction. The flat channel is marked as dark green dotted lines. (c) Overlay of microscope images of the non-moving single colloidal particles inside generic $\mathcal{P}$-tiles and of moving colloidal dumbbells inside the flat channel within one twisted square (top) and hexagonal (bottom) moiré Wigner Seitz cell. The colloidal dumbbells move through the flat channel (trajectory bright green line, flat channel dashed dark green line) with their long axes locked to the flat channel direction. (d) Microscope image of single colloidal particles inside generic $\mathcal{P}$-tiles and of colloidal dumbbells inside the flat channel. The dumbbells move through the flat channel of the square (top) and hexagonal (bottom) moiré Wigner Seitz cell while precessing with their long axes locked to the external field. The trajectory of one colloidal dumbbell is colored according to the period of the precessing external field (see magnified inset). The flat channel is located exactly at the marked trajectory. (e) Overlay of microscope images of the rotating but not sliding motion above the inclined dithered twisted square pattern. The flat channel is marked as dark green dotted lines. (f) Overlay of microscope images of the non rotating and not sliding motion above the inclined dithered twisted hexagonal pattern. The flat channel is marked as dark green dotted lines. The color of the frames in (b)–(f) correspond to the colors of the transport mode in panel (a). Videos of the different transport modes are provided with the Videos S1–S10 (ESI†).

We cover the moiré pattern with photoresist of thickness $t = 1$ μm, that separates colloids from the magnetic thin film such that only long wave length Fourier modes (wave length of the order of the $\mathcal{P}$-tile lattice constant or more) of the pattern magnetic field is relevant at the location of our colloids. The moiré pattern is surrounded by five computer-controlled coils, four of which create an external in plane field and the other produces an external magnetic field normal to the pattern. The moiré pattern with the coils is mounted to a microscope operating in reflection mode. The microscope itself is mounted to a support with the optical axis along the pattern normal. The patter normal is inclined $\hat{g}\cdot n = -\cos(\beta)$ with respect to gravity with an inclination angle of $\beta = \pi/9$. Paramagnetic colloidal particles of diameter 4.51 μm (Dynabeads M-450) are immersed in a drop of water placed on the pattern.

We apply a time dependent external field

$$H_{ext}(t) = R(\omega t) \cdot H_{ext}(0) \quad (2)$$

that rotates with angular frequency $\omega = \omega n$. Here $\omega = 1.5$ s$^{-1}$, and $R(\omega t)$ is a rotation matrix about the pattern normal $n$ and the external field $H_{ext}$ is tilted with an angle $\vartheta$ to the pattern normal ($H_{ext} \cdot n = \cos \vartheta$). The motion of the external field is thus a precession with angle $\vartheta$ around the moiré pattern normal.

Fig. 2 shows two dynamical phase diagrams of the motion above the inclined dithered twisted square and hexagonal patterns together with overlay of tracked video microscopy images of each mode of motion. We find five different transport modes.

The simplest mode occurs for low external magnetic field ($H_{ext} < 0.2$ kA m$^{-1}$) oriented close to the equatorial plane ($\vartheta > 50°$, blue triangles in Fig. 2a, microscopy images in Fig. 2b and Video S1 and S6, ESI†). The external field in this case precesses across the marginally stable points of the potential that will flatten the potential to valleys not only above the usual flat channels but also above the generic $\mathcal{P}$-tiles of the pattern. All colloidal particles, no matter where they are located, start to slide down the slope of the inclined plane under these circumstances. Based on the observation of multiple movies we see that in general the sliding direction of the colloidal particles in the flat channels follows an orientation of $\pm\pi/2n$ with respect to the inclination $(1 - nn) \cdot g$ direction of the pattern if the particles are far away from the corners of the flat channels, but follow the inclination direction $(1 - nn) \cdot g$ in the surroundings of the flat channel corners and in the generic $\mathcal{P}$-tiles. Colloidal dumbbells exhibit the same behavior. Additionally their long axis seems to correlate with the travel direction far away from the corners and seem to be less correlated while traveling in the corner surroundings.

In all other transport modes the external field will immobilize colloidal particles and dumbbells above the generic $\mathcal{P}$-tiles and the only remaining regions of mobility can be found above the flat channels or in the surroundings of the flat channel corners. At low external field, $H_{ext} < 0.2$ kA m$^{-1}$, and for precession angles $\vartheta < 50°$ dumbbells consisting of two colloidal particles slide through the flat channels with their long axis $d$ oriented along the channel (bright green triangles in Fig. 2a and microscopy images in Fig. 2c as well as in Videos S2 and S7, ESI†).

The trajectories follow the flat channel in the flat channel segments far from the corners, but they follow the in-plane gravity direction $(1 - nn) \cdot g$ in the surroundings of the flat channel corners. The orientation of the dumbbell becomes random in the corner surroundings and switches to the new direction of the flat channels once the dumbbell reaches the next flat channel segment (see also Fig. 3a and b). We call this phase the slither sliding phase. Since the external field is weak under these circumstances, the anisotropic moiré pattern potential torque dominates over putative torques due to the external magnetic field.

The long axis of the dumbbells lock to the external field if we use stronger external magnetic fields ($H_{ext} > 0.2$ kA m$^{-1}$). For precession angles above $\vartheta > 10°$, the long axis of the dumbbell precesses synchronously with the external field (Fig. 3c and d). The potential generated by the pattern for large precession angles remains weak enough such that gravity drives the precessing dumbbells through the flat channels (dark green circles in Fig. 2a and microscopy images in Fig. 2d as well as in Videos S3 and S8, ESI†). We call this phase the rotating and sliding phase.

The sliding stops for precession angles below $\vartheta < 50°$ as shown in the phase diagrams in Fig. 2a as orange circles. A microscopy image is depicted for half a moiré Wigner Seitz cell of the square pattern in Fig. 2e. Videos S4 and S9 (ESI†) record this mode for both patterns.

For even smaller precession angles $\vartheta < 10°$ (red triangles in Fig. 2a) the dumbbells also stop precessing. A microscope image Fig. 2f of part of a hexagonal moiré Wigner Seitz cell shows the static behavior. Videos S5 and S10 (ESI†) show a short movie of the static mode for both patterns. Presumably

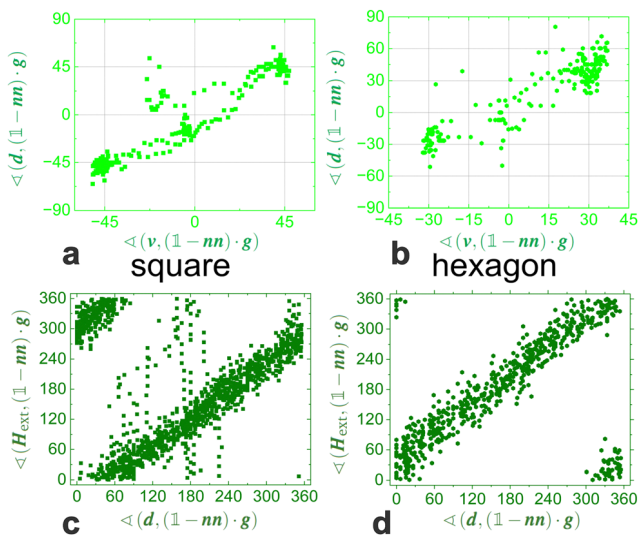

Fig. 3 Angle correlations (a) correlation between the velocity and director orientation in the slither sliding phase of the twisted square pattern. (b) Correlation between the velocity and director orientation in the slither sliding phase of the twisted hexagonal pattern. (c) Correlation between the director and external field orientation in the rotating and sliding phase of the twisted square pattern. (d) Correlation between the director and external field orientation in the rotating and sliding phase of the twisted hexagonal pattern.

the anisotropy in the colloidal potential dominates the behavior and suppresses any motion.

## 3. Discussion

In previous work[41] we have shown that magic non-generic transport behavior of colloids driven through a smooth twisted magnetic pattern occurs for magic angles due to the periodic nature of the corners of the flat channels. Under non-generic conditions each corner along a flat channel is different and in smooth magically twisted pattern one can avoid the occurrence of blockades at the corners. The transport on non-magically twisted patterns is predicted to stop at corners that block the transport. Smooth magically twisted systems therefore show pronounced non-generic transport behavior at the magic angles.

This is different for perturbed twisted systems: In intentionally heterostrained twisted bilayer graphene[56] the heterostrain can change the electronic flat bands. In experiments on a macroscopic scale we have shown[43] that disorder destroys the non-generic magic behavior such that the character of the transport is no longer decided at the corners but in the flat channel segments connecting the corners.

This is also true in the system studied here because the dithering, a specific form of disorder, of the twisted potential renders the flat channels rough. The roughness introduces obstacles that are harder to overcome and more frequent than those added under non-magical conditions at the corners. Therefore whether the potential is under non-generic twist angles or under magic conditions the transport behavior is decided inside the channels.

Single colloidal particles may travers potential barriers caused by the roughness of the potential inside the flat channels only for conditions that also mobilizes the single colloidal particles inside non-generic $\mathscr{P}$-tiles. Colloidal dumbbells carry double the weight and are long enough to pass over the roughness introduced by the dithering. This explains the slithering sliding of the dumbbells at small external field. Precession of the dumbbells under stronger external field introduces energetic fluctuations of the colloidal dumbbells that are strong enough to let the dumbbells pass over dithering obstacles. Hence the dumbbells perform a rotating sliding motion through the flat channels.

A smooth twisted colloidal potential could be produced by using two square or hexagonal patterns instead of the dithered single pattern. Colloidal particles would be placed between both patterns. However, it is difficult to visualize the colloids between both patterns.

## 4. Conclusions

In summary, twisted potentials are vulnerable to perturbations that usually destroys the non-generic magic behavior of the transport. However, other equally interesting transport modes that presumably persist whether the twist angle is magic or non-magic can be observed. Two distinct such modes: the slithering sliding and the rotating sliding have been characterized in this work.

## Author contributions

FF, DdlH, & TMF designed and performed the experiment, and wrote the manuscript with input from all the other authors. NCXS computed the dithered patterns. PK, MU & FS produced the magnetic film. SA, & ArE performed the fabrication of the micromagnetic metamorphic patterns within the magnetic thin film.

## Data availability

The data supporting this article have been included as part of the ESI.†

## Conflicts of interest

There are no conflicts to declare.

## Appendix

To achieve the observed phenomena we have to set the correct ratio of the relevant interactions. This is done the following way: The dipolar interaction between single colloidal particles scales with $H_{ext}^2$. We tried to work with external fields that produces singlets and dumbbells but not colloidal triplets, or quadruplets. Therefore we first chose an external field $H_{ext} < 1$ kA m$^{-1}$. The interaction potential of the particles with the pattern is proportional to $H_p^2$ for small external field and proportional to $\boldsymbol{H}_{ext} \cdot \boldsymbol{H}_p$ for large external fields. The inclination angle $\beta$ sets the driving force. There is a critical angle $\beta_c(H_{ext} = 1$ kA m$^{-1}) \approx 30°$. For $\beta > \beta_c$ the driving force is larger than the largest magnetic potential gradient such that the particles will start to slide in the generic positions of the moiré Wigner Seitz cell. We see that at large $H_{ext}$ increasing $\beta$ is similar to decreasing $H_{ext}$. If we do not consider the dithering roughness the phenomenon should only depend on the ratio $H_{ext}/\sin(\beta)$, however the dithering roughness makes things more complicated. We chose $\beta = 2/3\beta_c$ to obtain sliding in flat channels only. The frequency of rotation is such that the dumbbell can follow the external field when it is large but cannot follow when the external field is small. There is a non universal pattern potential that holds dumbbells oriented along a flat channel at vanishing or small external field. Reducing the precession frequency has no significant effect. There is a large cut off frequency where the response of the dumbbells is no longer synchronous to the field even without the magnetic pattern. We used frequencies well below this cutoff frequency.

## Acknowledgements

T. M. F. and D. d. l. H. acknowledge funding by the Deutsche Forschungsgemeinschaft (DFG, German Research Foundation)

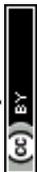

under project number 531559581. P. K., M. U., and F. S. acknowledge financial support from the National Science Centre Poland through the OPUS funding (Grant No. 2019/33/B/ST5/02013). S. A. acknowledges funding by a PhD scholarship of Kassel university.

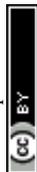

## Notes and references

1 Y. Cao, V. Fatemi, A. Demir, S. Fang, S. L. Tomarken, J. Y. Luo, J. D. Sanchez-Yamagishi, K. Watanabe, T. Taniguchi, E. Kaxiras, R. C. Ashoori and P. Jarillo-Herrero, *Nature*, 2018, **556**, 80–84.
2 R. Bistritzer and A. H. MacDonald, *Proc. Natl. Acad. Sci. U. S. A.*, 2011, **108**, 12233–12237.
3 K. P. Nuckolls, M. Oh, D. Wong, B. Lian, K. Watanabe, T. Taniguchi, B. A. Bernevig and A. Yazdani, *Nature*, 2020, **588**, 610–615.
4 E. Suárez Morell, J. D. Correa, P. Vargas, M. Pacheco and Z. Barticevic, *Phys. Rev. B: Condens. Matter Mater. Phys.*, 2010, **82**, 121407.
5 S. Carr, D. Massatt, S. Fang, P. Cazeaux, M. Luskin and E. Kaxiras, *Phys. Rev. B: Condens. Matter Mater. Phys.*, 2017, **95**, 075420.
6 J. M. B. Lopes dos Santos, N. M. R. Peres and A. H. Castro Neto, *Phys. Rev. Lett.*, 2007, **99**, 256802.
7 C. Shen, Y. Chu, Q. Wu, N. Li, S. Wang, Y. Zhao, J. Tang, J. Liu, J. Tian, K. Watanabe, T. Taniguchi, R. Yang, Z. Y. Meng, D. Shi, O. V. Yazyev and G. Zhang, *Nat. Phys.*, 2020, **16**, 520–525.
8 K. Liu, L. Zhang, T. Cao, C. Jin, D. Qiu, Q. Zhou, A. Zettl, P. Yang, S. G. Louie and F. Wang, *Nat. Commun.*, 2014, **5**, 4966.
9 H. Isobe, N. F. Q. Yuan and L. Fu, *Phys. Rev. X*, 2018, **8**, 041041.
10 L. Balents, C. R. Dean, D. K. Efetov and A. F. Young, *Nat. Phys.*, 2020, **16**, 725–733.
11 Y. Saito, J. Ge, K. Watanabe, T. Taniguchi and A. F. Young, *Nat. Phys.*, 2020, **16**, 926–930.
12 L. H. C. M. Nunes and C. M. Smith, *Phys. Rev. B*, 2020, **101**, 224514.
13 P. Stepanov, I. Das, X. Lu, A. Fahimniya, K. Watanabe, T. Taniguchi, F. H. L. Koppens, J. Lischner, L. Levitov and D. K. Efetov, *Nature*, 2020, **583**, 375–378.
14 Y. Cao, V. Fatemi, S. Fang, K. Watanabe, T. Taniguchi, E. Kaxiras and P. Jarillo-Herrero, *Nature*, 2018, **556**, 43–50.
15 M. Yankowitz, S. Chen, H. Polshyn, Y. Zhang, K. Watanabe, T. Taniguchi, D. Graf, A. F. Young and C. R. Dean, *Science*, 2019, **363**, 1059–1064.
16 A. L. Sharpe, E. J. Fox, A. W. Barnard, J. Finney, K. Watanabe, T. Taniguchi, M. A. Kastner and D. Goldhaber-Gordon, *Science*, 2019, **365**, 605–608.
17 X. Lu, P. Stepanov, W. Yang, M. Xie, M. A. Aamir, I. Das, C. Urgell, K. Watanabe, T. Taniguchi, G. Zhang, A. Bachtold, A. H. MacDonald and D. K. Efetov, *Nature*, 2019, **574**, 653–657.
18 Y. Zhao, M. A. Belkin and A. Alù, *Nat. Commun.*, 2012, **3**, 870.
19 S. S. Sunku, G. X. Ni, B. Y. Jiang, H. Yoo, A. Sternbach, A. S. McLeod, T. Stauber, L. Xiong, T. Taniguchi, K. Watanabe, P. Kim, M. M. Fogler and D. N. Basov, *Science*, 2018, **362**, 1153–1156.
20 G. Hu, Q. Ou, G. Si, Y. Wu, J. Wu, Z. Dai, A. Krasnok, Y. Mazor, Q. Zhang, Q. Bao, C.-W. Qiu and A. Alù, *Nature*, 2020, **582**, 209–213.
21 P. Wang, Y. Zheng, X. Chen, C. Huang, Y. V. Kartashov, L. Torner, V. V. Konotop and F. Ye, *Nature*, 2020, **577**, 42–46.
22 G. Hu, C.-W. Qiu and A. Alù, *Opt. Mater. Express*, 2021, **11**, 1377–1382.
23 A. I. Cocemasov, D. L. Nika and A. A. Balandin, *Phys. Rev. B: Condens. Matter Mater. Phys.*, 2013, **88**, 035428.
24 G. Ma, M. Xiao and C. T. Chan, *Nat. Rev. Phys.*, 2019, **1**, 281–294.
25 S. M. Gardezi, H. Pirie, S. Carr, W. Dorrell and J. E. Hoffman, *2d Matter.*, 2021, **8**, 031002.
26 Y. Deng, M. Oudich, N. J. Gerard, J. Ji, M. Lu and Y. Jing, *Phys. Rev. B*, 2020, **102**, 180304.
27 D. Yao, L. Ye, Z. Fu, Q. Wang, H. He, J. Lu, W. Deng, X. Huang, M. Ke and Z. Liu, *Phys. Rev. Lett.*, 2024, **132**, 266602.
28 W. Li, C. J. O. Reichhardt, B. Jankó and C. Reichhardt, *Phys. Rev. B*, 2021, **104**, 024504.
29 A. M. E. B. Rossi, J. Bugase and T. M. Fischer, *Europhys. Lett.*, 2017, **119**, 40001.
30 A. M. E. B. Rossi, J. Bugase, T. Lachner, A. Ernst, D. de las Heras and T. M. Fischer, *Soft Matter*, 2019, **15**, 8543–8551.
31 J. Loehr, M. Loenne, A. Ernst, D. de las Heras and T. M. Fischer, *Nat. Commun.*, 2016, **7**, 11745.
32 D. de las Heras, J. Loehr, M. Loenne and T. M. Fischer, *New J. Phys.*, 2016, **18**, 105009.
33 J. Loehr, D. de las Heras, M. Loenne, J. Bugase, A. Jarosz, M. Urbaniak, F. Stobiecki, A. Tomita, R. Huhnstock, I. Koch, A. Ehresmann, D. Holzinger and T. M. Fischer, *Soft Matter*, 2017, **13**, 5044–5075.
34 J. Loehr, D. de las Heras, A. Jarosz, M. Urbaniak, F. Stobiecki, A. Tomita, R. Huhnstock, I. Koch, A. Ehresmann, D. Holzinger and T. M. Fischer, *Commun. Phys.*, 2018, **1**, 4.
35 H. Massana-Cid, A. Ernst, D. de las Heras, A. Jarosz, M. Urbaniak, F. Stobiecki, A. Tomita, R. Huhnstock, I. Koch, A. Ehresmann, D. Holzinger and T. M. Fischer, *Soft Matter*, 2019, **15**, 1539–1550.
36 M. Mirzaee-Kakhki, A. Ernst, D. de las Heras, M. Urbaniak, F. Stobiecki, J. Gördes, M. Reginka, A. Ehresmann and T. M. Fischer, *Nat. Commun.*, 2020, **11**, 4670.
37 M. Mirzaee-Kakhki, A. Ernst, D. de las Heras, M. Urbaniak, F. Stobiecki, A. Tomita, R. Huhnstock, I. Koch, J. Gördes, A. Ehresmann, D. Holzinger, M. Reginka and T. M. Fischer, *Soft Matter*, 2020, **16**, 1594–1598.
38 M. Mirzaee-Kakhki, A. Ernst, D. de las Heras, M. Urbaniak, F. Stobiecki, A. Tomita, R. Huhnstock, I. Koch, A. Ehresmann, D. Holzinger and T. M. Fischer, *Soft Matter*, 2021, **17**, 1663–1674.

39 N. C. X. Stuhlmüller, F. Farrokhzad, P. Kuświk, F. Stobiecki, M. Urbaniak, S. Akhundzada, A. Ehresmann, T. M. Fischer and D. de las Heras, *Nat. Commun.*, 2023, **14**, 7517.
40 J. Elschner, F. Farrokhzad, P. Kuswik, M. Urbaniak, F. Stobiecki, S. Akhundzada, A. Ehresmann, D. de las Heras and T. M. Fischer, *Nat. Commun.*, 2024, **15**, 5735.
41 N. C. X. Stuhlmüller, T. M. Fischer and D. de las Heras, *Commun. Phys.*, 2022, **5**, 48.
42 N. C. X. Stuhlmüller, T. M. Fischer and D. de las Heras, *New J. Phys.*, 2024, **26**, 023056.
43 A. M. E. B. Rossi, A. Ernst, M. Dörfler and T. M. Fischer, *Commun. Phys.*, 2024, **7**, 24.
44 G. Costantini and F. Marchesoni, *EPL*, 1999, **48**, 491.
45 M. Tinkham, J. U. Free, C. N. Lau and N. Markovic, *Phys. Rev. B: Condens. Matter Mater. Phys.*, 2003, **68**, 134515.
46 C. Guarcello, D. Valenti, A. Carollo and B. Spagnolo, *J. Stat. Mech.: Theor. Exp*, 2016, 054012.
47 S. A. Tatarkova, W. Sibbett and K. Dholakia, *Phys. Rev. Lett.*, 2003, **91**, 038101.
48 K. Lindenberg, J. M. Sancho, A. M. Lacasta and I. M. Sokolov, *Phys. Rev. Lett.*, 2007, **98**, 020602.
49 P. Siegle, I. Goychuk and P. Hänggi, *Phys. Rev. Lett.*, 2010, **105**, 100602.
50 C. Reichhardt and C. J. O. Reichhardt, *Phys. Rev. B: Condens. Matter Mater. Phys.*, 2015, **92**, 224432.
51 M. Urbaniak, P. Kuświk, Z. Kurant, M. Tekielak, D. Engel, D. Lengemann, B. Szymański, M. Schmidt, J. Aleksiejew, A. Maziewski, A. Ehresmann and F. Stobiecki, *Phys. Rev. Lett.*, 2010, **105**, 067202.
52 P. Kuświk, A. Ehresmann, M. Tekielak, B. Szymański, I. Sveklo, P. Mazalski, D. Engel, J. Kisielewski, D. Lengemann, M. Urbaniak, C. Schmidt, A. Maziewski and F. Stobiecki, *Nanotechnology*, 2011, **22**, 095302.
53 D. Lengemann, D. Engel and A. Ehresmann, *Rev. Sci. Instrum.*, 2012, **83**, 053303.
54 F. J. Maier, T. Lachner, A. Vilfan, T. O. Tasci, K. B. Neeves, D. W. M. Marr and T. M. Fischer, *Soft Matter*, 2016, **12**, 9314–9320.
55 S. Barthmann and T. M. Fischer, *J. Phys. Commun.*, 2021, **5**, 085003.
56 F. Mesple, A. Missaoui, T. Cea, L. Huder, F. Guinea, G. Trambly de Laissardière, C. Chapelier and V. T. Renard, *Phys. Rev. Lett.*, 2021, **127**, 126405.